\documentclass[12pt]{article}
\pdfoutput=1
\usepackage{amsmath}

\usepackage{amssymb}
\usepackage{graphicx}
\usepackage{color}
\newlength{\abstractwidth}
\usepackage[T1]{fontenc}
\usepackage[utf8]{inputenc}
\usepackage{graphicx}% Include figure files
\usepackage{bm}% bold math
\usepackage{xcolor}
\newcommand\eqn{\addtocounter{equation}{1}\tag{\theequation}}

\def\t{\tau}
\def\tt{\tilde{\tau}}
\def\ts{\tilde{\sigma}}
\def\fg{\mathfrak{g}}
\def\J{\mathcal{J}}
\def\K{\mathcal{K}}
\def\N{\mathcal{N}}
\DeclareMathOperator{\Tr}{Tr}

\begin{document}

\begin{titlepage}
\bigskip
\bigskip\bigskip 
\bigskip

\begin{center}
%\centerline
{\Large \bf{Thermodynamics and Many Body Chaos for generalized large $q$ SYK models}}
 \bigskip
%\centerline
{\Large \bf { }} 
\bigskip
\bigskip
\end{center}

\begin{center}

\bf {Jiaqi Jiang and Zhenbin Yang}
\bigskip \rm
\bigskip \rm

\bigskip
\bigskip
Jadwin Hall, Princeton University,  Princeton, NJ 08540, USA

\end{center}
\bigskip\bigskip
\begin{abstract}
\medskip
\noindent
This paper considers a type of generalized large $q$ SYK models which include multi-body interactions between Majorana fermions.
 We derive an effective action in the limit of large $N$ and large $q$ (with ${~q^2\over N} $ small), and find a universal expression for thermodynamical quantities.
We also consider the chaos exponent using the retarded kernel method and find an efficient way to calculate the Lyapunov exponent for generalized large $q$ SYK models numerically.
\end{abstract}
\bigskip \bigskip \bigskip 
 
\end{titlepage}

\tableofcontents

\section{Introduction}
The Sachdev-Ye-Kitaev model \cite{Sachdev:1992fk,Kitaev:2015video,Kitaev:2017awl,Maldacena:2016hyu} is a quantum mechanics model describing Near Extremal Black Holes \cite{Jackiw:1984je,Teitelboim:1983ux}. It is also the first example demonstrating NAdS$_2$/NCFT$_1$ holographic duality \cite{Almheiri:2014cka,Maldacena:2016upp,Jensen:2016pah,Engelsoy:2016xyb}.
The standard SYK model contains $N$ Majorana fermions with random $q$-local interactions. 
We consider a variant of this model by including various types of all-to-all interactions and solve the theory in the double-scaling limit \cite{2019arXiv190311115S, 2017JHEP...05..118C,2014MPAG...17..441E,Berkooz:2018jqr,Qi:2018bje} where we take $q$ and $N$ to infinity and keep $\frac{q^2}{N}$ fixed and small.
We show that the large $N$ effective action \eqref{N_eff} describes a two dimensional scalar field with a general potential.
Using the effective action, we derive its themodynamic relation and the Lyapunov exponent of OTOC (out-of-time-ordered correlator).
We find that the OTOC is given by the Lorentzian propagator of the scalar field and therefore is controlled by its potential.
In the generalized large $q$ SYK model we are considering, the chaos exponents correspond to the energies of the bound states and we will show that in such models there exists a unique Lyapunov exponent.
\\

This work can be used to understand the qualitative behavior of Lyapunov exponent under relevant deformations.  
Our generic expectation about the Lyapunov exponent is that it is governed by IR modes of the theory. This means irrelevant deformations will not induce any change of the Lyapunov exponent as well relevant deformations will become important as we lower the temperature. And when the relevant deformation grows, the dynamics will be dominated by the lowest dimensional operator, e.g if we add a mass term the theory will be gapped and there is no chaos behavior \cite{Bi:2017yvx,Chen:2017dav}; if the lowest dimensional operator is four fermion interaction or higher, the theory flows to the standard SYK model with maximal Lyapunov exponent.  It is the behavior of the transition along the RG flow that we want to address.\\

The paper is organized as follows: 
in section two, we set up the analytic investigation of the generalized large $q$ SYK model and derive the double-scaling effective action; in sections three and four we derive its thermodynamic relation and study the Lyapunov exponent; 
in section five, we explore some concrete examples including large $q$ and $2q$ model, and a scaling model where the interactions contain all $q$-fermion interactions with a particular distribution of the coupling strength;
in section six, we show that the Lyapunov exponent is unique.

\section{Generalized large $q$ SYK model}
SYK is a zero dimensional quantum mechanics model of $N$ Majorana fermions $\chi_i$ with all to all $q$ fermion interactions:
\begin{equation}\label{H_disc}
H=\sum_{1\leq i_1<...<i_q\leq N} (i)^{\frac{q}{2}}J_{i_1...i_q}\chi_{i_1}...\chi_{i_q}
\end{equation}
The random interaction strength $J_{i_1...i_q}$ satisfies a gaussian distribution: $\langle J_{i_1...i_q}J_{k_1...k_q}\rangle=\frac{J^2 (q-1)!}{N^{q-1}}\delta_{k_1...k_q}^{i_1...i_q}$.

There are several important features of this model in the limit of $N$ going to infinity: First, Feynman diagrams are dominated by melonic diagrams;
Second, the theory is exactly solvable if we further assume $q$ is large;
Third, when $q>2$, the model develops conformal symmetry at IR;
Fourth, the conformal symmetry is slightly broken leaving only a SL(2,R) subgroup unbroken, and the pseudo-Golstone boson is controlled by the Schwarzian action;
Last, the unbroken SL(2,R) symmetry gives rise to the maximally chaotic behavior of SYK at IR \cite{2019arXiv190412820L,2019arXiv190412819S}.

\begin{figure}[h]
\centering
\includegraphics[scale=.3]{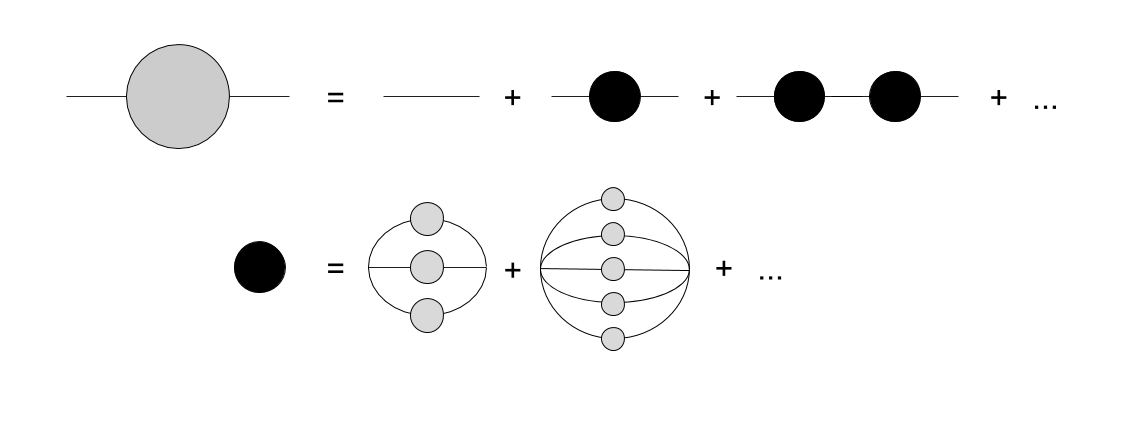}
\caption{The diagrammatic representation of the Schwinger-Dyson equations. The grey circle represents the full two point function and the black circle represents the one particle irreducible contributions.}
\label{SD diagrams}
\end{figure}

\begin{figure}[ht]
\centering
\includegraphics[scale=.4]{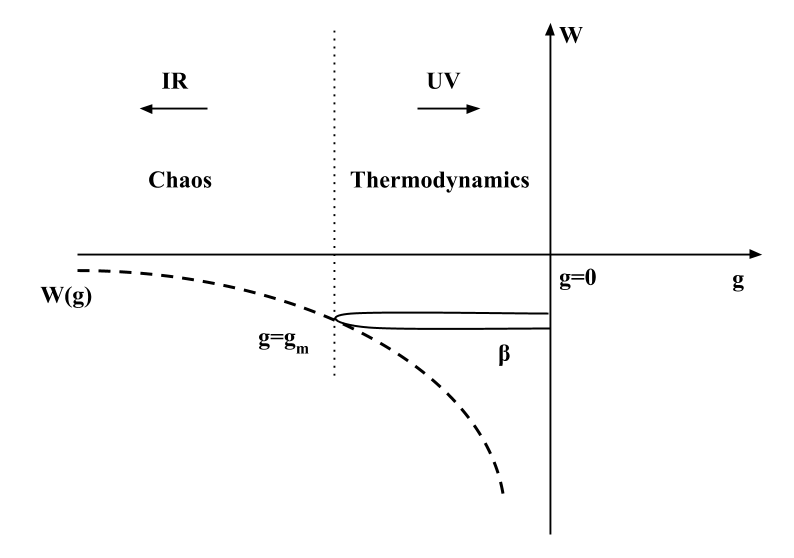}
\caption{The figure depicts the scattering of a particle from the potential $W(g)$. The total time it takes for the particle to return to the origin defines $\beta$. Increasing $\beta$ corresponds to scattering the particle with a higher energy and the particle penetrates deeper into the $g<0$ region. The thermodynamic quantities are determined by the potential in the region $g_m\leq g\leq 0$ while the Lyapunov exponent is determined by the potential in the region $g\leq g_m$.}
\label{Potential}
\end{figure}

If one is only interested in the low energy behavior of this theory then only the lowest dimensional operator matters so it can be classified to be either gapped (when the lowest dimension operator is two-local) or maximally chaotic (when the lowest dimension operator is four-local or higher).  However, since here we want to talk about the Lyapunov exponent at all temperature scales, different types of interactions will indeed change its behavior. On that account we extend our Hamiltonian to include different types of interactions denoted by $q_i ~ (1\leq i\leq K)$:
\begin{equation}
H=\sum_{q=q_1}^{q_K}\sum_{1\leq i_1<...<i_q\leq N} (i)^{\frac{q}{2}} J_{i_1...i_q}\chi_{i_1}...\chi_{i_q}
\end{equation}
We are taking $J_{i_1...i_q}$ to be independent gaussian disorder variables with zero mean and variance $J_q^2 \frac{(q-1)!}{N^{q-1}}$. 
In the limit of large $N$, we can use melon diagrams (figure \ref{SD diagrams}) to write down the Schwinger-Dyson equation in Euclidean time $\tau$:
\begin{equation}\label{eqn-SDequation}
\frac{1}{G(\omega)}=-i\omega-\Sigma(\omega),\quad \Sigma(\tau)=\sum_{q=q_1}^{q_K} J^2_q\left[G(\tau)\right]^{q-1}
\end{equation}
where we define the disordered average two point function in a thermal state $G(\tau_1-\tau_2)=\sum_i\frac{1}{N}\langle T\chi_i(\tau_1)\chi_i(\tau_2)\rangle_{\beta} $.
$\langle T\chi_i(\tau_1)\chi_i(\tau_2)\rangle_{\beta}$ is defined as
\begin{equation}
\langle T\chi_i(\tau_1)\chi_i(\tau_2)\rangle_{\beta}=\theta(\tau_1-\tau_2)\Tr\left[\rho(\beta) \chi_i(\tau_1)\chi_i(\tau_2)\right]-\theta(\tau_2-\tau_1)\Tr\left[\rho(\beta) \chi_i(\tau_2)\chi_i(\tau_1)\right],
\end{equation}
with $\rho(\beta)=e^{-\beta H}$ being the density matrix and we have used the time translation invariance to simplify the two point function.
$\Sigma(\tau)$ is the self energy and is represented by the black bulb in figure \ref{SD diagrams}.
The self energy is a sum of different products of two point function determined by our interaction.
Allowing some of the $J_{q_i}$'s to be zero, we can assume $\{q_i\}$ is a sequence with increment 2 from $q_1$ to $q_K$.
Clearly, from the SD equation, the low energy limit is only determined by $q_1$  in the summation of self energy.
To solve the SD equation at all temperature scales analytically, we can take the large $q$ limit. To be more precise, we consider the case that all the $q_i$ scale to infinity at the same rate:
\begin{equation}
	q_i=\alpha_i q;~~~~{1\over q}\ll\alpha_1\ll1\ll\alpha_K\ll q;~~~\delta\alpha=\alpha_{i+1}-\alpha_i={2\over q};~~
\end{equation}
Considering $J^2(\alpha)={q\over 2} J^2_{\alpha q}$, one can write down an integral expression of the SD equation:
\begin{equation}
	\frac{1}{G(\omega)}=-i\omega-\Sigma(\omega),~~~~~~~~
\Sigma(\tau)=\lim_{q\rightarrow\infty}\sum_i \delta \alpha J^2(\alpha_i)G(\tau)^{\alpha_i q-1}=\int_{0}^{\infty} d\alpha J^2(\alpha)G(\tau)^{\alpha q-1}.
\end{equation}
where the lower bound $\alpha_1$ and upper bound $\alpha_K$ are taken to be 0 and $\infty$ respectively. 
This SD equation can be formally obtained by the following symbolic Hamiltonian:
\begin{equation}\label{LargeQHamiltonian}
H = \int_{0}^{\infty} d\alpha (i)^{\frac{\alpha q}{2}} \sum_{M} J_{M}(\alpha)\chi_{M}^{\alpha q}
\end{equation}
where $(i)^{\frac{\alpha q}{2}} \sum_{M} J_{M}(\alpha)\chi_{M}^{\alpha q}$ stands for a SYK type $\alpha q$-local interactions.
$M=\lbrace i_1...i_{\alpha q}\rbrace$ enumerates all possible $\alpha q$ fermions. And the interactions strength satisfies:
\begin{equation}\label{JJ_correlator}
\langle J_{M}(\alpha)J_{M'}(\beta)\rangle = \frac{J^2(\alpha)\Gamma(\alpha q)}{N^{\alpha q-1}}\delta(\alpha-\beta)\delta_{M,M'}
\end{equation}
We call this model (\ref{LargeQHamiltonian}) the generalized large $q$ SYK model.\footnote{Strictly speaking, this Hamiltonian is only defined if the $J(\alpha)$ is centered at even integer values of $\alpha q$.  In large q limit, as we have shown above, one can just  treat it as a positive continuous function of $\alpha$.}
One can also try to keep some of the $q_i$ finite, which corresponds to deforming the Hamiltonian with a finite $q$ deformation. One such example is to deform the large $q$ SYK with mass term, which has a low energy interpretation of a double trace deformation of JT gravity \cite{Maldacena:2018lmt,Maldacena:2017axo,2017JHEP...12..151G,Chen:2019qqe}.
We can use the following ansatz to solve the SD equations \footnote{Higher order in $\frac{1}{q}$ was considered in \cite{Tarnopolsky:2018env}.}
\begin{align}
&G(\tau)=\frac{1}{2}\text{sgn}(\tau)\left[1+\frac{1}{q}g(\tau)+\dots\right],\label{sd_ansatz1}\\
&\Sigma(\tau)=\int_0^\infty  d\alpha J^2(\alpha)2^{1-\alpha q}\text{sgn}(\tau)e^{\alpha g(\tau)}(1+\dots).\label{sd_ansatz2}
\end{align}
The SD equations imply that
\begin{equation}\label{def_ug}
g''(\tau)=U(g)\equiv 2\int_0^\infty d\alpha \mathcal{J}^2(\alpha)e^{\alpha g(\tau)}
\end{equation}
where $\mathcal{J}^2(\alpha)=q2^{1-\alpha q}J^2(\alpha)$ and we define $U(g)$ such that the SD equations become Newton's equation for a particle under a classical force $U(g)$.  The KMS condition demands that the particle bounces back to its original location after time $\beta$, i.e. $g(0)=g(\beta)=0$ (See figure \ref{Potential}). We also see that $U(g)$ can be very general since it is a Laplace transformation of an arbitrary positive function $\mathcal{J}^2(\alpha)$. If we further define the potential $W(g)=-\int_{-\infty}^g d\tilde g U(\tilde g)$, then it follows that
\begin{equation}\label{eq_gw}
g'(\tau)=-\sqrt{2\left[W(g_m)-W(g)\right]},\quad \tau(g)=
\begin{cases}
\int_{g}^0  \frac{dg}{\sqrt{2[W(g_m)-W(g)]}},\quad &\tau<\frac{\beta}{2}\\
\\
\beta-\int_{g}^0  \frac{dg}{\sqrt{2[W(g_m)-W(g)]}},\quad &\tau >\frac{\beta}{2}\\
\end{cases}
\end{equation}
where $g_m$ is the location at which the particle bounces ($g_m\leq g(\tau)\leq 0$) back and it is related to $\beta$ by:
\begin{equation}\label{eq_bw}
\beta=\sqrt{2}\int_{g_m}^0  \frac{dg}{\sqrt{W(g_m)-W(g)}} 
\end{equation}
From the definition of $U(g)$, the potential $W(g)$ has the property that all its derivatives are negative and in particular it is monotonic and always less than zero. Therefore, when $g_m$ becomes more and more negative, the temperature of the system approaches to zero. So the system flows from UV to IR as we move leftwards along the $g$-axis as shown in Figure (\ref{Potential}).

The large $q$ Schwinger-Dyson equation can also be derived from the large $N$ effective action obtained from the original fermion path integral using the replica method.  In the standard large $q$ SYK model, one arrives at a Liouville field theory \cite{Maldacena:2018lmt}. Using the same derivation with some small modifications, our effective action becomes:
\begin{equation}\label{N_eff}
\boxed{S_E=-S_0+\frac{N}{8q^2}\int_0^{\beta} d\t_1 \int_0^{\beta} d\t_2 \left[\frac{1}{2}\partial_{\t_1}g(\t_1,\t_2)\partial_{\t_2}g(\t_1,\t_2)+ W(g(\t_1,\t_2))\right],}
\end{equation}
where $S_0=\frac{N}{2}\log 2$ is the entropy of N free Majorana fermions. The derivation of the effective action is included in appendix \ref{App:effective action}. 
The $g$ field is defined in equation (\ref{sd_ansatz1}). It is symmetric with respect to $\t_1$ and $\t_2$ as dictated by fermion statistics. Also from the KMS condition, it is periodic with inverse temperature $\beta$.  Finally, since the theory is free at UV, it vanishes at coincident points: $g(\t,\t)=0$.
We can also define the following coordinates
\begin{equation}\label{kin_coord}
\tt=\t_1+\t_2,\qquad \ts=\t_1-\t_2
\end{equation}
which can be considered as describing the kinematic space. In terms of the kinematic space coordinates, the action \eqref{N_eff} describes the most general type of scalar field theory, though we need to keep in mind that the possible forms of the potential $W(g)$ are restricted.
From this effective action, we can calculate the two point function $G(\t_1,\t_2)$ to leading order $\frac{1}{q}$ by simply taking its expectation value:
\begin{equation}\label{eq:relaton_G_g}
	G(\t_1,\t_2)=\frac{1}{2}\text{sgn}(\tau_{12})+\frac{\text{sgn}(\tau_{12})}{2q}\langle g(\tau_1,\tau_2)\rangle
\end{equation}
where $\langle g(\tau_1,\tau_2)\rangle$ represents doing functional integral over $g$ with the effective action (\ref{N_eff}).
In the large $N$ limit, the expectation value is determined from its classical solution and is the same as \eqref{eq_gw} from the SD equation after using the translation symmetry.

 Below, we will use this action and its classical solution to study the thermodynamics and chaos behavior of the generalized large $q$ SYK model.

\section{Thermodynamic quantities}

In the leading large $N$ approximation, the free energy can be obtained by simply evaluating the on-shell action (\ref{N_eff}) of its classical solution. In particular, we will study the classical solution which has the translation symmetry so that $g(\tau_1,\tau_2)=g(\tau_{12})$. The on-shell action for such classical solution \eqref{eq_gw} then becomes (subtracted by $S_0$):
\begin{align}
\beta F=\frac{N\beta}{8q^2}\int_0^\beta d\t\left[-\frac{1}{2}\partial_\t g(\tau)\partial_\t g(\tau)+W(g(\t))\right]
\end{align}
It will be convenient to define the constant $\N=\frac{N}{8q^2}$. We can use \eqref{eq_gw} to get rid of the kinetic term and arrive at the following expression for free energy:
\begin{equation}\label{free_en_1}
F=\N\int_0^\beta d\t\left[-\frac{1}{2}\partial_\t g(\tau)\partial_\t g(\tau)+W(g(\t))\right]=\N\beta W(g_m)-2\sqrt{2}\N\int^0_{g_m}dg\sqrt{W(g_m)-W(g)}
\end{equation}
where $g_m$ as we have discussed in the previous section is the locus where the particle bounces back. Since $g_m$ is uniquely determined by the temperature $\beta$, we shall treat $g_m$ as the thermodynamic variable.
We can calculate the energy using the following argument: if we consider $\J^2(\alpha)=\J^2f(\alpha)$ where $\J$ is a constant that defines the scale, then $\beta$ and $\J$ should appear together in the expression. The only $\J$ dependence in \eqref{N_eff} is from the potential $W(g)$ and this gives us
\begin{equation}
\J\partial_\J (-\beta F)=-2\N\int d\t_1 d\t_2 W(g(\t_1,\t_2))=-\beta E
\end{equation}
Therefore, we can also express the energy $E$ in terms of the potential $W(g)$ and the thermodynamic variable $g_m$ as:
\begin{equation}\label{en_1}
E=2\N\beta W(g_m)-2\sqrt{2}\N\int^0_{g_m}dg\sqrt{W(g_m)-W(g)}
\end{equation}
We find that the expression of the free energy $F$ and the expression of the energy $E$ only differs by $\N \beta W(g_m)+T S_0$ which is equal to the product of temperature and entropy. Therefore, all the thermodynamic quantities can be determined from the potential $W(g)$ and the variable $g_m$ in the generalized large $q$ SYK model:\footnote{It is interesting that the the final result has an explicit dependence on the potential. A similar behavior happens for 2d dilaton gravity where the thermodynamic relation only depends on a similar quantity called prepotential \cite{Maldacena:2019cbz}}
\begin{equation}\label{eq_F_S}
F=-\N\beta W(g_m)+E-\frac{S_0}{\beta}, \quad S=S_0+\N\beta^2 W(g_m)
\end{equation}
It is not obvious from the expressions \eqref{eq_F_S} that the energy $E$ and the entropy $S$ would satisfy the thermodynamic relation. However, we can check explicitly that these expressions indeed satisfy the thermodynamic relation $\partial_{\beta} E=\frac{1}{\beta}\partial_{\beta} S$. Because the thermodynamic variable $g_m$ is uniquely determined by $\beta$, we can equivalently check the relation $\partial_{g_m} E=\frac{1}{\beta}\partial_{g_m} S$. The explicit calculations give
\begin{align}
\frac{1}{\beta}\partial_{g_m} S=\N\left(2W(g_m)\frac{d \beta(g_m)}{dg_m}+\beta \partial_gW(g_m)\right)
=2\N\frac{d}{dg_m}\left(\beta W(g_m)\right)-\N\beta \partial_gW(g_m)
\end{align}
\begin{align}
\partial_{g_m}E&=2\N\frac{d}{dg_m}\left(\beta W(g_m)\right)-2\sqrt{2}\N\frac{d}{dg_m}\left(\int^0_{g_m}dg\sqrt{W(g_m)-W(g)}\right)\nonumber \\
&=2\N\frac{d}{dg_m}\left(\beta W(g_m)\right)-\N\beta \partial_gW(g_m)
\end{align}
We see that our expressions for the thermodynamic quantities of the generalized large $q$ SYK model do satisfy the thermodynamic relation.

\section{Chaos exponent}

In this section we study the chaos behavior of the generalized large $q$ SYK model. We need to consider the following out-of-time-ordered correlation function (OTOC) in Lorentizian time \cite{Kitaev:2015video}
\begin{equation}\label{def_otoc}
F(t_1,t_2)=\frac{1}{N^2}\sum_{i,j=1}^N\text{Tr}\left[y\chi_i(t_1)y\chi_j(0)y\chi_i(t_2)y\chi_j(0)\right], \quad y\equiv \rho(\beta)^{1/4}.
\end{equation}
The fermions in \eqref{def_otoc} are separated by a quarter of the thermal circle.  Instead of calculating $F(t_1,t_2)$ directly, we could obtain it by the analytic continuation of the Euclidean correlator $\langle G(\t_1,\t_2)G(0,0)\rangle_\beta$. Since $G(\tau_1,\tau_2)$ is related to $g(\tau_1,\tau_2)$ by \eqref{eq:relaton_G_g}, we need to compute the Euclidean correlator $\langle g(\t_1,\t_2)g(0,0)\rangle_\beta$ using the large $N$ effective action \eqref{N_eff}. It is more convenient if we change to the kinematic space coordinates \eqref{kin_coord} and analytically continue back to Lorentzian signature. The resulted Lorentzian action is 
\begin{equation}
i S_{M}=i \mathcal{N} \int dt d\sigma \left[\frac{1}{4}\left(\partial_t g(t,\sigma)\partial_t g(t,\sigma)-\partial_\sigma g(t,\sigma)\partial_\sigma g(t,\sigma)\right)-\frac{1}{2}W(g)\right]
\end{equation}
where $t$ and $\sigma$ are the analytic continuations of the kinematic space coordinates $\tt$ and $\ts$ defined in \eqref{kin_coord}. In the large $N$ limit, the leading connected piece in $\langle g(t,\sigma)g(0,0)\rangle_\beta$ ($t=t_1+t_2$ and $\sigma=t_1-t_2$) will be given by the two-point function of the fluctuation $\fg(t,\sigma)$  around the classical solution $g_c(\sigma)$: $\langle\fg(t,\sigma)\fg(0,0)\rangle_\beta$. The quadratic effective action for $\fg(t,\sigma)$ is 
\begin{equation}
i S_{\fg\fg}=i\mathcal{N}\int dt d\sigma \frac{1}{4}\left[\partial_t \fg(t,\sigma)\partial_t \fg(t,\sigma)-\partial_\sigma \fg(t,\sigma)\partial_\sigma \fg(t,\sigma)-\partial^2_{g_c} W(g_c(\sigma))\fg^2(t,\sigma)\right]
\end{equation}
The two-point function $K(t,\sigma)=\langle\fg(t,\sigma)\fg(0,0)\rangle_\beta$ is just the propagator of $\fg$, which satisfies the differential equation
\begin{equation}\label{eq:chaos_1}
\left[-\partial_t^2+\partial_\sigma^2-\partial^2_{g_c} W(g_c(\sigma))\right]K(t,\sigma)=\frac{2 i}{\mathcal{N}}\delta(t)\delta(\sigma)
\end{equation}
Since we want to study the exponential growth of $K(t,\sigma)$ at late time, we will use the ansatz $K(t,\sigma)=e^{\frac{\lambda t}{2}}f_\lambda(\sigma)$ as guaranteed by the translation symmetry of $t$ in \eqref{eq:chaos_1} and look in the regime $t\gg 1$. After plugging the ansatz into \eqref{eq:chaos_1}, we obtain the following differential equation for $f_\lambda(\sigma)$:
\begin{equation}\label{eq:chaos_2}
\left(\frac{\lambda^2}{4}-\partial_\sigma^2\right)f_\lambda(\sigma)=-\partial^2_{g_c} W(g_c(\sigma))f_\lambda(\sigma)
\end{equation} 
which is same as the Schr$\ddot{o}$dinger equation for a particle moving in the potential $\partial^2_{g_c} W(g_c(\sigma))$. Thus, the problem of finding the Lyapunov exponent for the generalized large $q$ SYK model becomes a quantum mechanical problem of finding the spectrum of the bound states for a particle in the $\partial^2_{g_c} W(g_c(\sigma))$ potential. Although the form of potential $W$ as a function of $g$ is simple, the form of the classical solution $g_c(\sigma)$ is usually complicated. Therefore, it is more useful if we can express \eqref{eq:chaos_2} purely in terms of variable $g$. We can obtain a relation between the two variables $g$ and $\sigma$ by analytically continuing \eqref{eq_gw}:
\begin{equation}
\tau = \frac{\beta}{2}+i\sigma = -\int^g_{0} d\tilde{g}\frac{1}{\sqrt{2(W(g_m)-W(\tilde{g}))}}=\frac{\beta}{2}-i\int^g_{g_m} d\tilde{g}\frac{1}{\sqrt{2(W(\tilde{g})-W(g_m))}}
\end{equation}
Here the variable $g$ is in the range $(-\infty,g_m)$, so the particular branch we have chosen corresponds to $\sigma\in(0,\infty)$. For $\sigma\in(-\infty,0)$, we need to choose the other branch to do the analytic continuation. Here, we will just work with the branch corresponding to $\sigma>0$ so we can express $\sigma$ in terms of $g$ by the following equation:
\begin{equation}\label{eq:sigma_g}
\sigma = \int^{g_m}_g d\tilde{g}\frac{1}{\sqrt{2(W(\tilde{g})-W(g_m))}}
\end{equation}
with $g\in(-\infty,g_m)$. Recall that when we calculate the various thermodynamic quantities in the previous section, we only need the information about the potential $W(g)$ in the region $g\in(g_m,0)$ while here we see that the chaos behavior is completely determined by the part of the potential $W(g)$ in the region $g\in(-\infty,g_m)$ in figure \ref{Potential}.
This manifests the fact that the chaos behavior is controlled by the IR degrees of freedom of the system.
Now we can use \eqref{eq:sigma_g} to write the differential equation \eqref{eq:chaos_2} purely in terms of $g$ variable with $g$ in the range of $(-\infty, g_m)$ as
\begin{equation}\label{lyap_W}
2\sqrt{W(g)-W(g_m)}\partial_g\left(\sqrt{W(g)-W(g_m)}\partial_g f_{\lambda}(g)\right)=\left(\frac{\lambda^2}{4}+\partial_g^2W(g)\right)f_{\lambda}(g)
\end{equation}
which we can use to calculate numerically the Lyapunov exponent without solving \eqref{eq_gw} directly. The boundary condition for the ground state wavefunction is $f_\lambda'(g_m)=0$.

\section{Results of specific models}

\subsection{Large $q$ and $2q$ model}
We consider the model which corresponds to the SYK model with a large $q$ and $2q$ interactions. Specifically, the Hamiltonian for the model is
\begin{equation}
H=(i)^{\frac{q}{2}}\sum_{1\leq i_1<\dots<i_q\leq N} j_{i_1\dots i_q}\chi_{i_1}\dots\chi_{i_q}+(i)^q\sum_{1\leq i_1<\dots<i_{2q}\leq N} k_{i_1\dots i_{2q}}\chi_{i_1}\dots\chi_{i_{2q}}
\end{equation}
\begin{equation}
\langle j^2_{i_1\dots i_{q}}\rangle=\frac{2^{q-1}}{q}\frac{\J^2(q-1)!}{N^{q-1}},\quad \langle k^2_{i_1\dots i_{2q}}\rangle=\frac{2^{2q-1}}{2q}\frac{\K^2(2q-1)!}{N^{2q-1}}
\end{equation}

Using our formalism, this corresponds to a generalized large $q$ SYK model with the following $U(g)$:
\begin{equation}\label{geq_q_2q}
g''(\t)=U(g)=2\J^2e^g+\K^2e^{2g}
\end{equation}
The corresponding potential $W(g)$ is then $W(g)=-2\J^2 e^g-\frac{1}{2}\K^2 e^{2g}$. We can express the inverse temperature $\beta$ in terms of the thermodynamic variable $g_m$ defined in previous section by using \eqref{eq_bw}. We then obtain the following relation:
\begin{equation}\label{eq:q_2q_beta}
\beta = \frac{2 e^{-\frac{g_m}{2}}\Theta}{\sqrt{4\J^2+e^{g_m}\K^2}}
\end{equation}
where the $\Theta$ is defined as
\begin{equation}\label{eq:q_2q_theta}
\cos\Theta = \frac{(-2+4e^{g_m})\J^2+e^{2g_m}\K^2}{2\J^2+e^{g_m}\K^2}.
\end{equation}
For fixed $\J$ and $\K$, we see that as $g_m$ goes from $0$ to $-\infty$, the angle $\Theta$ ranges from $0$ to $\pi$ and $\beta$ increases from $0$ to $\infty$. Since $\K$ appears in both \eqref{eq:q_2q_beta} and \eqref{eq:q_2q_theta} with $e^{g_m}$ prefactor, at low temperature ($g_m\ll 0$) $\K$ is strongly suppressed, which is expected because at IR the $q$-fermion interaction dominates.  
The entropy for this model can be calculated using \eqref{eq_F_S} and we get
\begin{equation}
S=S_0-2\N\Theta^2
\end{equation}
At zero temperature, the correction becomes $-\frac{\pi^2}{4q^2}$ which is identical to the standard large $q$ SYK model as expected.

We can numerically calculate the Lyapunov exponent for this model using \eqref{lyap_W}. On the other hand, we can actually solve \eqref{geq_q_2q} analytically and then use perturbation theory to calculate the Lyapunov exponent at the low temperature regime where the $q$-fermion interaction dominates and at the high temperature regime where the $2q$-fermion interaction dominates. The details of the perturbative calculations are included in the Appendix \ref{App:pert q and 2q}. We show in figure \ref{fig:q_2q} the numerical result for the Lyapunov exponent at different temperature scales and compare it with the analytic results from the perturbation theory.
The red (dotted) curve shows the behavior of the Lyapunov exponent at the temperature where the 2$q$-fermion interaction dominates, and the perturbation calculation shows that adding the relevant interaction decreases the chaos exponent until the relevant perturbation dominates which is described by the green (dashed) curve.
At lower temperature the theory is described by standard SYK with single $q$-fermion interactions.

\begin{figure}[ht]
\centering
\includegraphics[scale=.6]{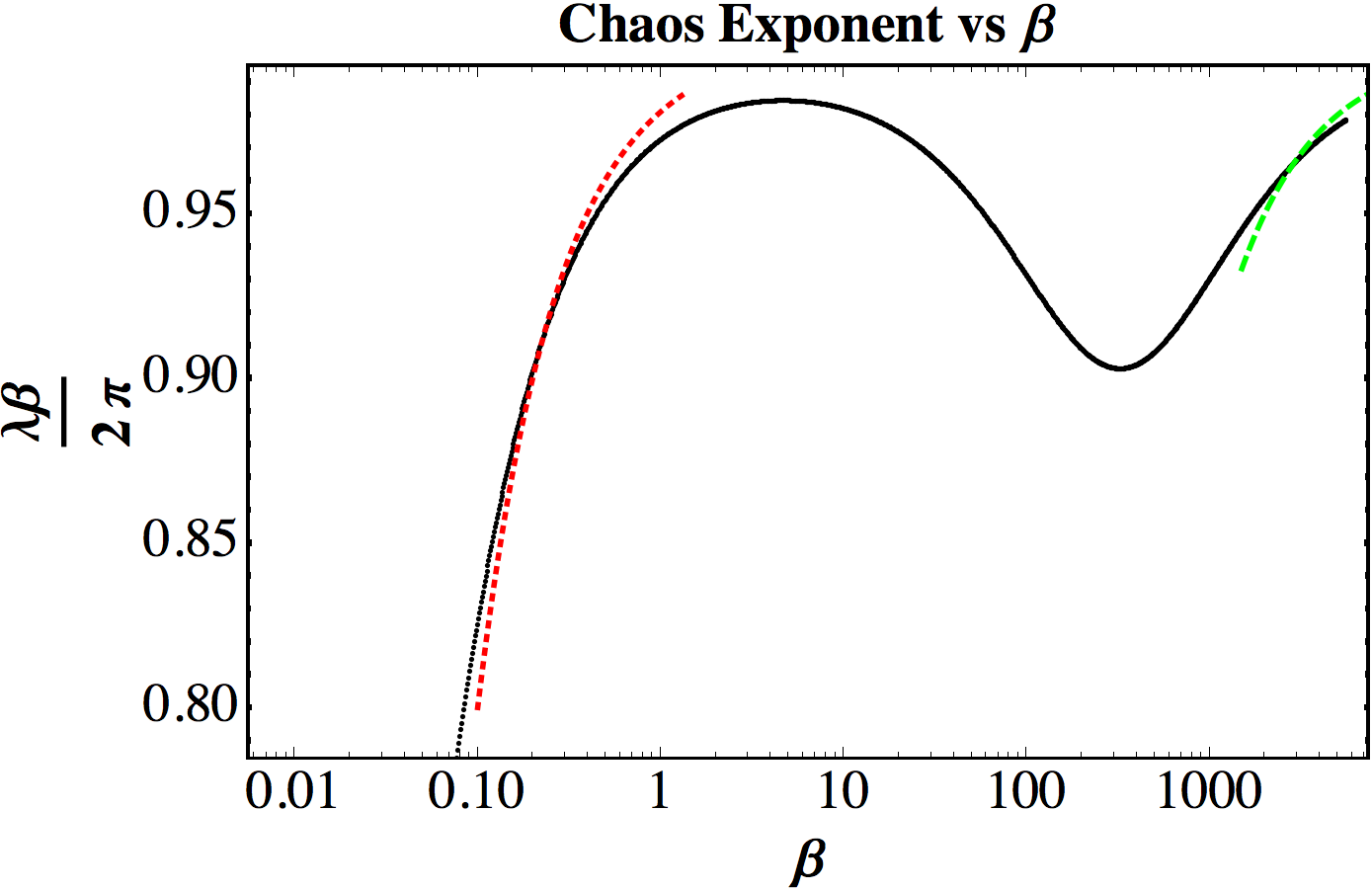}
\caption{The figure shows the log plot of the Lyapunov exponent against $\beta$ for the large $q$ and $2q$ model with $\J=1$ and $\K=100$. The red (dotted) curve shows the result from the perturbative calculation when $\beta\K\gg 1$ and $(\beta\J)^2\ll \beta\K$, while the green (dashed) curve shows the result from the perturbative calculation when $\beta\J\gg 1$ and $(\beta\J)^2\gg \beta\K$.}
\label{fig:q_2q}
\end{figure}

\subsection{Scaling model}
In this section, we consider the model with the couplings $\mathcal{J}^2(\alpha)=\mathcal{J}^2\alpha^n$ 
which we will refer as the scaling model for the reason shown afterwards. This model corresponds to the following $U(g)$ and $W(g)$:\footnote{This is an approximation only applicable in the large $q$ limit where the lower cutoff should be order of $\frac{1}{q}$.  There could also be an upper bound for the number of fermion interactions, in that situation our result applies to the intermediate temperature region where those high dimension operators become negligible.}
\begin{equation}
U(g)=2\int^\infty_0 d\alpha \J^2 \alpha^n e^{\alpha g}=\frac{2\J^2\Gamma(n+1)}{(-g)^{n+1}},\quad W(g)=-\frac{2\J^2\Gamma(n)}{(-g)^n}
\end{equation}
The thermodynamic quantities can be calculated using \eqref{en_1} and \eqref{eq_F_S}:
\begin{equation}
S = S_0-\frac{2\N \beta^2\J^2\Gamma(n)}{(-g_m)^n},\quad E=\frac{8\N\beta\J^2\Gamma(n)}{(n-2)(-g_m)^n}	
\end{equation}
The relation between $g_m$ and $\beta$ is given by \eqref{eq_bw} as 
\begin{equation}\label{sc_betaj}
\beta\J=\sqrt{\frac{\pi}{\Gamma(n)}}\frac{\Gamma(\frac{1}{n}+\frac{1}{2})}{\Gamma(\frac{1}{n})}(-g_m)^{\frac{n+2}{2}}
\end{equation}
So we can express $S$ and $\beta E$ in terms of $\beta\J$ as
\begin{equation}
S=S_0-2\N(\beta\J)^{\frac{4}{n+2}}C_n,\quad \beta E=\frac{8\N}{n-2}(\beta\J)^{\frac{4}{n+2}}C_n
\end{equation}
with $C_n$ a constant coefficient that only depends on $n$. We see that this model has the interesting feature that the entropy $S-S_0$ and the energy $E$ scale with $\beta$. It is for this reason that we refer the model as the scaling model.

\begin{figure}[ht]
\centering
\includegraphics[scale=.6]{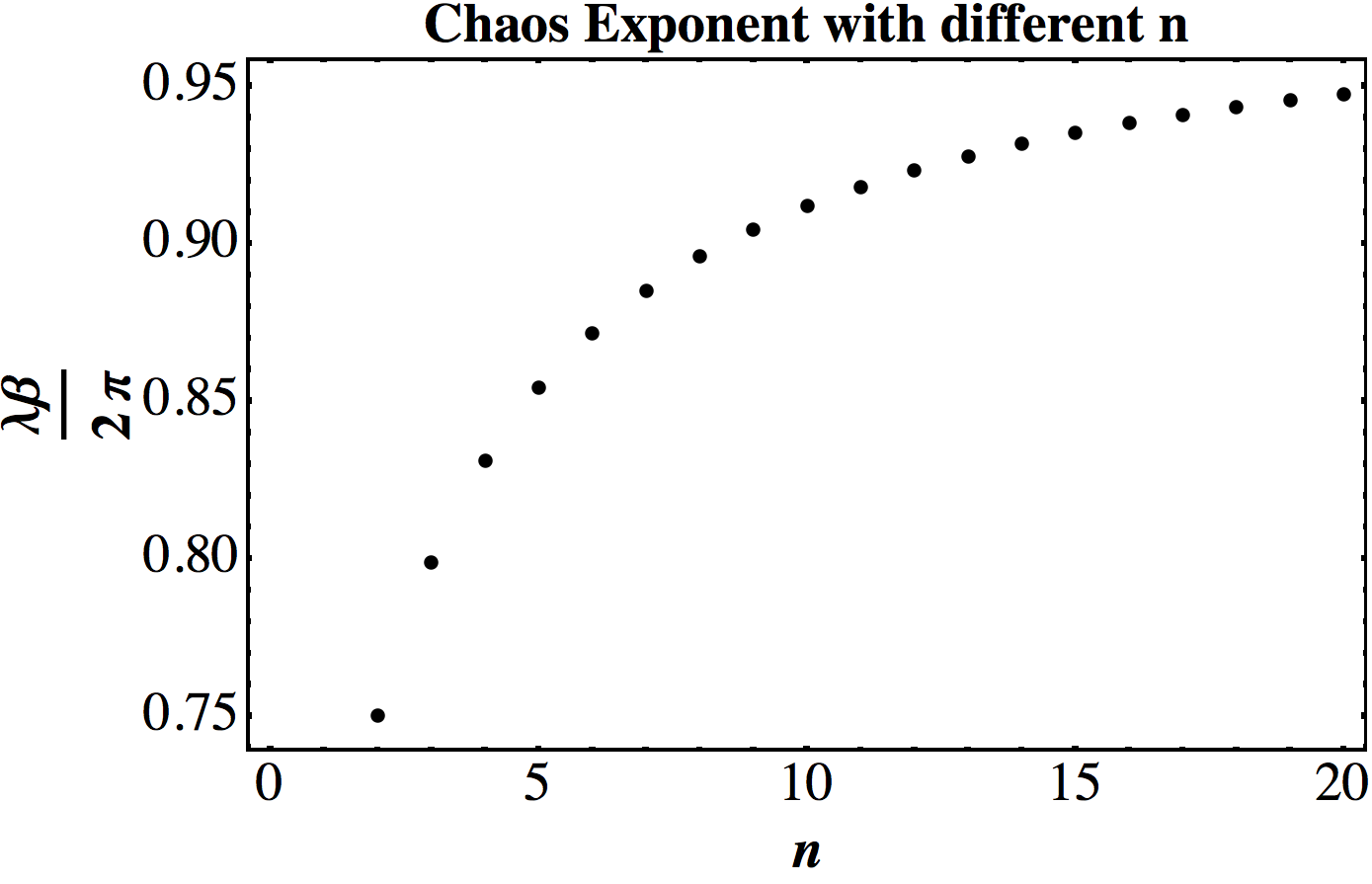}
\caption{The figure shows the Lyapunov exponent for the scaling model with different values of $n$. The rescaled Lyapunov exponent approaches the chaos bound as $n$ increases.}
\label{chaos_scaling_model}
\end{figure}

The Lyapunov exponent can be calculated numerically using \eqref{lyap_W}. If we introduce the new variable $x\in [1,\infty)$ defined by $g=xg_m$ and use the relation \eqref{sc_betaj}, then the equation \eqref{lyap_W} becomes
\begin{equation}
\left(2-\frac{2}{x^n}\right)f''(x)+\frac{n}{x^{n+1}}f'(x)+\frac{n(n+1)}{x^{n+2}}f(x)=\left(\frac{\lambda\beta}{2\pi}\right)^2 A_n f(x)
\end{equation}
with $A_n=\frac{\pi\Gamma^2(\frac{1}{n})}{2\Gamma^2(\frac{1}{n}+\frac{1}{2})}$. We see that the rescaled chaos exponent $\frac{\lambda\beta}{2\pi}$ does not change with temperature in this model, a feature absent in other SYK models. We plot in figure \ref{chaos_scaling_model} the rescaled chaos exponent against the power $n$ and we observe that the rescaled chaos exponent approaches to the chaos bound as $n$ increases.

\section{Eigenvalue structure of the chaos exponent equation}
As we have shown in the previous section, the problem of calculating the chaos exponent in the generalized large $q$ SYK model with the potential $W(g)$ is equivalent to the quantum mechanics problem of finding the energy spectrum of the bound states for a particle moving in the potential $\partial^2_g W(g)$. 
From the definition \eqref{def_ug} of $U(g)$ and $U(g)=-\partial_g W(g)$, we see that the possible form of $W(g)$ is quite restricted in the generalized large $q$ SYK model. Specifically, $W(g)$ has to be the Laplace transformation of a negative distribution and thus has the following properties:
\begin{enumerate}
\item $W(g)$ goes to a constant as $g\to -\infty$. This constant is arbitrary so we can always set it to be zero.
\item Any number of derivatives of $W$ is always negative. In particular, $W(g)$ is monotonically decreasing.
\item $W(g)$ is well defined for $g\in(-\infty, 0)$.
\end{enumerate}
As a result, the potential $\partial^2_gW(g)$ will always have a bound state. This implies that there is always an exponential growth at the late time for OTOC in the generalized large $q$ SYK model. 

It is natural to ask if there is any subleading exponential growth in the late time OTOC for the model.\footnote{We thank Y.Gu and D.Stanford for helpful discussion on this.} In terms of the equivalent quantum mechanics problem, this translates to the question if there exists any other bound states besides the ground state. Here we shall argue that there is no such bound states, so no subleading chaos growth in the generalized large $q$ SYK model. Although it is difficult to solve the bound state spectrum directly for general potential $\partial^2_g W(g)$, we notice that there always exits a scattering state with $\lambda_L=0$. The wavefunction of this state is given by
\begin{equation}\label{eq:lambda0_wavefunction}
f(\sigma)=g'(\sigma), \text{ or } f(g)=\sqrt{W(g)-W(g_m)}
\end{equation}
We can verify this directly by plugging into the chaos exponent equation \eqref{eq:chaos_1} and recall that $g''(t)=-U(g)=\partial_gW(g)$. 
Such a mode with zero chaos exponent has to exist because of energy conservation and we see explicitly that the eigenfunction is generated by taking a time derivative. Furthermore, this scattering state has only one zero point at $g=g_m$ since $W(g)$ monotonically decreases. By the node theorem from quantum mechanics, this should be the first excited state and therefore the spectrum of the system consists of only one single bound state, which means a unique Lyapunov exponent.

\section{Conclusion}
In this paper we studied the generalized large $q$ SYK models. We derived the expressions for the thermodynamic quantities such as energy (\ref{en_1}) and entropy (\ref{eq_F_S}) and wrote down the general equation to calculate the chaos exponent (\ref{lyap_W}) in such models.
We pointed out that the equation (\ref{lyap_W}) is convenient to do numerical calculations and analyzed its eigenvalue structures. In particular our analysis showed that there exists only one Lyapunov exponent in the generalized large $q$ SYK models and we expect this is a general feature for ladder diagram dominated models.
We studied two particular models: the first is the large $q$ and $2q$ model where the chaos exponent displays initial decrease under relevant deformation; the second is the scaling model where the chaos exponent is a constant ratio of the maximum value at all temperatures. 

\vspace{0.75cm}
{\bf Acknowledgments } 

We thank Y.Gu, J.Maldacena, D.Stanford, J.Zhang and W.Zhao for helpful discussions. 
Z.Y is supported by Charlotte Elizabeth Procter Fellowship from Princeton University.

\appendix
\section{Derivation of the effective action}\label{App:effective action}
In this section, we compute the free energy which is equivalent to the effective action by using the replica trick \cite{2017JHEP...02..093G}
\begin{equation}
\beta \overline{F}=-\overline{\ln Z}=-\lim_{M\to 0}\frac{\ln \overline{Z^M}}{M}
\end{equation}
where the bar indicates averaging over the disorder and $Z^M$ is the partition function of $M$ copies of the system. Specifically, we have
\begin{align*}
\overline{Z^M}=\int \mathcal{D}J_I(\alpha)&\int\mathcal{D}\chi_i^a P[J_I]\\
&\exp\left[-\sum_a \int d\t\left( 
\frac{1}{2}\sum_i \chi_i^a\partial_\t \chi_i^a 
+ \int^\infty_0  \frac{d\alpha(i)^{\frac{\alpha q}{2}}}{\Gamma(\alpha q+1)}\sum_I J_I(\alpha)\chi^{a,\alpha q}_I
\right) \right] \eqn
\end{align*}
where $a$ is the replica index, $a\in \{1,\dots,M\}$, $i\in \{1,\dots, N\}$, and $I$ in $J_I(\alpha)$ is the collective index representing $i_1\dots i_{\alpha q}$, and $P[J_I]$ is the probability distribution for $J_I(\alpha)$ which gives \eqref{JJ_correlator}. We can regard $J_I$ as a dynamical field with propagator \eqref{JJ_correlator} and integrate it out. The result is
\begin{align*}\label{appendix_eq1}
\overline{Z^M}=\int \mathcal{D}\chi_i^a\exp & \left[-\sum_{a,i}\frac{1}{2}\int d\t\chi_i^a\partial_\t \chi_i^a \right.\\
+&\left.\frac{1}{4}\int_0^\infty d\alpha \frac{\J^2(\alpha) N}{\alpha q^2}\sum_{a,b}\int d\t_1 d\t_2 \left(\frac{2}{N}\sum_{i}\chi_i^a(\t_1)\chi_i^b(\t_2)\right)^{\alpha q}\right] \eqn
\end{align*}
Next we introduce the collective fields
\begin{equation}
G^{ab}(\t_1,\t_2)=\frac{1}{N}\sum_i \chi_i^a(\t_1)\chi_i^b(\t_2)
\end{equation}
and insert into \eqref{appendix_eq1} the following delta function 
\begin{equation}
\int d\Sigma^{ab}(\t_1,\t_2)\exp\left[-\frac{N}{2}\Sigma^{ab}(\t_1,\t_2)\left(G^{ab}(\t_1,\t_2)-\frac{1}{N}\sum_i\chi_i^a(\t_1)\chi_i^b(\t_2)\right)\right]
\end{equation}
This leads to
\begin{align*}
\overline{Z^M}=&\int \mathcal{D}\Sigma \mathcal{D} G\int \mathcal{D}\chi_i^a\exp\left[-\frac{1}{2}\sum_{a,i}\int d\t\chi_i^a\partial_\t \chi_i^a
+\frac{1}{2}\sum_{a,b,i}\int d\t_1 d\t_2\Sigma^{ab}(\t_1,\t_2)\chi_i^a(\t_1)\chi_i^b(\t_2)\right.\\
&\left.-\frac{N}{2}\sum_{a,b}\int d\t_1 d\t_2\left(\Sigma^{ab}(\t_1,\t_2)G^{ab}(\t_1,\t_2)-\int_0^\infty d\alpha\frac{\J^2(\alpha)}{2\alpha q^2}\left(2G^{ab}(\t_1,\t_2)\right)^{\alpha q}\right)\right] \eqn
\end{align*}
Now we can integrate out the fermions and arrive at
\begin{align*}
\overline{Z^M}=&\int \mathcal{D}\Sigma \mathcal{D}G \exp  \left\{N\sum_{a,b}\left[\Tr\log(\delta^{ab}\partial_\tau-\Sigma^{ab})\right.\right.\\
-&\left.\left.\frac{1}{2}\int d\t_1 d\t_2\left(\Sigma^{ab}(\t_1,\t_2)G^{ab}(\t_1,\t_2)-\frac{1}{2q^2}\int_0^\infty d\alpha\frac{\J^2(\alpha)}{\alpha}\left(2G^{ab}(\t_1,\t_2)\right)^{\alpha q} \right)\right]\right\}\eqn
\end{align*}
Finally, we shall assume a replica symmetric saddle point so that $G^{ab}(\t_1,\t_2)=\delta^{ab}G(\t_1,\t_2)$ and $\Sigma^{ab}(\t_1,\t_2)=\delta^{ab}\Sigma(\t_1,\t_2)$, then we have 
\begin{equation}
\overline{Z^M}=\int \mathcal{D}\Sigma \mathcal{D}G \exp\left(-M S_E\right)
\end{equation}
where the large N effective action $S_E$ is
\begin{equation}\label{l_largeN}
-\frac{S_E}{N}=\frac{1}{2}\text{Tr}\log(\partial_\tau-\Sigma)-\frac{1}{2}\int d\t_1 d\t_2\left[\Sigma(\t_1,\t_2)G(\t_1,\t_2)-\frac{1}{2q^2}\int_0^\infty d\alpha \frac{\J^2(\alpha)}{\alpha}[2G(\t_1,\t_2)]^{\alpha q}\right]
\end{equation}
Using \eqref{sd_ansatz1} and \eqref{sd_ansatz2}, the determinant term can be expanded in large $q$ limit as
\begin{equation}
\text{Tr}\log(\partial_{\t}-\Sigma)=\text{Tr}\log(G_0^{-1})-\text{Tr}(G_0*\Sigma)-\frac{1}{2}\text{Tr}(G_0*\Sigma*G_0*\Sigma)+\dots
\end{equation}
with $G_0(\t_1,\t_2)=\frac{1}{2}\text{sgn}(\t_1-\t_2)$. The $\text{Tr}\log(G_0^{-1})$ term gives the entropy of free fermions while the $\text{Tr}(G_0*\Sigma)$ term vanishes. Ignoring the constant piece, we have to order $q^{-2}$:
\begin{align*}\label{q^2_eff}
\frac{S_E}{N}\simeq &\frac{1}{4}\text{Tr}(G_0*\Sigma*G_0*\Sigma)+\frac{1}{2q}\int d\t_1 d\t_2 \Sigma(\t_1,\t_2)G_0(\t_1,\t_2)g(\t_1,\t_2)\\
&-\frac{1}{4q^2}\int d\t_1 d\t_2 \int_0^\infty d\alpha \frac{\J^2(\alpha)}{\alpha}e^{\alpha g(\t_1,\t_2)} \eqn
\end{align*}
If we define
\begin{equation}
\Phi(\t_1,\t_2)=[G_0*\Sigma](\t_1,\t_2)
\end{equation}
Then it follows that
\begin{equation}
\Sigma(\t_1,\t_2)=\partial_{\t_1}\Phi(\t_1,\t_2)
\end{equation}
and \eqref{q^2_eff} becomes
\begin{align*}
\frac{S_E}{N}\simeq &\frac{1}{4}\text{Tr}(\Phi*\Phi)+\frac{1}{2q}\int d\t_1 d\t_2 \partial_{\t_1}\Phi(\t_1,\t_2)G_0(\t_1,\t_2)g(\t_1,\t_2)\\
&-\frac{1}{4q^2}\int d\t_1 d\t_2 \int_0^\infty d\alpha \frac{\J^2(\alpha)}{\alpha}e^{\alpha g(\t_1,\t_2)} \eqn
\end{align*}
After integrating out $\Phi$, we obtain (after subtracting the constant piece)
\begin{align*}
\frac{S_E}{N}&=\frac{1}{4q^2}\int d\t_1 d\t_2\partial_{\t_1}(G_0(\t_1,\t_2)g(\t_1,\t_2))\partial_{\t_2}(G_0(\t_1,\t_2)g(\t_1,\t_2))\\
&\quad -\frac{1}{4q^2}\int d\t_1 d\t_2 \int_0^\infty d\alpha \frac{\J^2(\alpha)}{\alpha}e^{\alpha g(\t_1,\t_2)}\\
&=\frac{1}{16q^2}\int d\t_1 d\t_2\partial_{\t_1}g(\t_1,\t_2)\partial_{\t_2}g(\t_1,\t_2)-\frac{1}{4q^2}\int d\t_1 d\t_2 \int_0^\infty d\alpha \frac{\J^2(\alpha)}{\alpha}e^{\alpha g(\t_1,\t_2)}\\
&=\frac{1}{16q^2}\int d\t_1 d\t_2\partial_{\t_1}g(\t_1,\t_2)\partial_{\t_2}g(\t_1,\t_2)+\frac{1}{8q^2}\int d\t_1 d\t_2 W(g(\t_1,\t_2))
\eqn\label{eff_act1}
\end{align*}
To arrive at the first line, we use the property that $g(\tau_1,\tau_2)=g(\tau_2,\tau_1)$ from the definition of $G(\tau_1,\tau_2)$ and the last line follows from the definition of $U(g)$ and $W(g)$.

\section{Thermodynamics of the standard large $q$ SYK model}
In the large $q$ limit of the standard SYK model \cite{Maldacena:2016hyu}, we have $U(g)=2\J^2 e^g$ and $W(g)=-2\J^2 e^g$. Then it follows from \eqref{eq_bw} that
\begin{equation}
\beta\J = 2 \exp\left(-\frac{g_m}{2}\right)\tan^{-1}\left(\sqrt{e^{-g_m}-1}\right)
\end{equation}
If we define $\nu$ such that $\tan(\frac{\pi\nu}{2})=\sqrt{e^{-g_m}-1}$, then we have the relations
\begin{equation}\label{def_nu}
\beta\J=\frac{\pi\nu}{\cos(\frac{\pi\nu}{2})},\qquad 2\J^2\cos^2\left(\frac{\pi\nu}{2}\right)=-W(g_m)
\end{equation}
Using \eqref{def_nu} and \eqref{eq_F_S}, we have
\begin{equation}
S=S_0-\frac{N}{4q^2}\pi^2\nu^2
\end{equation}
We can use \eqref{def_nu} to obtain the low-temperature expansion of $\nu$ as
\begin{equation}
\nu=1-\frac{2}{\beta\J}+o(\tfrac{1}{(\beta\J)^2})
\end{equation}
which reproduces the $-\frac{\pi^2}{4q^2}$ term of the zero-temperature entropy of the standard large-$q$ model result.

\section{Perturbative calculations in large $q$ and $2q$ model}\label{App:pert q and 2q}
The equation \eqref{geq_q_2q} can be solved analytically and we have the solution
\begin{equation}\label{q_2q_sol}
e^{g(t)}=\frac{2\nu^2}{\sqrt{\J^4+\nu^2\K^2}\cos(2\nu t-\nu\beta)+\J^2},\qquad \text{with } \frac{2\nu^2-\J^2}{\sqrt{\J^4+\nu^2\K^2}}=\cos(\nu\beta)
\end{equation}
Since $\nu$ is dimensionful, if we introduce $\nu=\tfrac{\omega}{\beta}$ with $\omega\sim o(1)$, then the equation for $\nu$ becomes
\begin{equation}
\frac{2\omega^2-(\beta\J)^2}{\sqrt{(\beta\J)^4+\omega^2(\beta\K)^2}}=\cos(\omega)
\end{equation}
from which we see that the transition from two conformal points occurs at the temperature $(\beta\J)^2\sim\beta\K$.

If we plug the solution \eqref{q_2q_sol} into \eqref{eq:chaos_2} and define $x=2\nu \sigma$, together with $\J^2=A\cos\theta$ and $\nu\K=A\sin\theta$ where $\theta\in(0,\frac{\pi}{2})$, we then have
\begin{equation}\label{q_2q_lbd_eqn}
-\partial_x^2f_\lambda(x)-\frac{\cos\theta}{\cosh x+\cos\theta}f_\lambda(x)-\frac{2\sin^2\theta}{(\cosh x+\cos\theta)^2}f_\lambda(x)=-(\frac{\lambda}{4\nu})^2f_\lambda(x)
\end{equation}
Although we are not able to solve \eqref{q_2q_lbd_eqn}, we can consider doing perturbation around the two conformal points and evaluate the correction to the Lyapunov exponent. 

We first consider the case that $\epsilon=\frac{(\beta\J)^2}{\beta\K}\ll 1$ and $\beta\K\gg 1$, then we have
\begin{equation}
\omega=\frac{\pi}{2}+\frac{2}{\pi}\epsilon-\frac{\pi}{\beta\K}+o(\tfrac{1}{(\beta\K)^2}),\quad \theta=\frac{\pi}{2}-\frac{2}{\pi}\epsilon + o(\tfrac{1}{(\beta\K)^2})
\end{equation}
The left-hand-side of perturbed equation \eqref{q_2q_lbd_eqn} becomes
\begin{equation}
-\partial_x^2f_\lambda(x)-\frac{2}{\cosh^2x}f_\lambda(x)-\epsilon\left(\frac{1}{\cosh x}-\frac{4}{\cosh^3 x}\right)f_\lambda(x)
\end{equation}
Using the unperturbed ground state solution $f_\lambda(x)=\frac{1}{\sqrt{2}\cosh x}$, we get 
\begin{equation}
\lambda =4\nu(1-\frac{1}{2}\epsilon)=\frac{2\pi}{\beta}\left[1-\frac{2}{\beta\K}+(\frac{1}{2}-\frac{4}{\pi^2})\frac{(\beta\J)^2}{\beta\K}\right]+o(\tfrac{1}{(\beta\K)^2})
\end{equation}

We can also consider the case that $\frac{1}{\beta\J}\ll \gamma=\frac{\beta\K}{(\beta\J)^2}\ll 1$, then we have
\begin{equation}
\theta=\omega\gamma+o(\gamma^2),\quad \omega=\pi-\pi\gamma+o(\tfrac{1}{\beta\J})
\end{equation}
The l.h.s of the perturbed equation becomes
\begin{equation}
-\partial_x^2f_\lambda(x)-\frac{1}{2\cosh^2\tfrac{x}{2}}f_\lambda(x)-\frac{\gamma^2\pi^2}{2}\frac{4-\cosh x}{1+\cosh^2 x}f_\lambda(x)
\end{equation}
However, in this case we see that the perturbation starts at $o(\gamma^2)$, which implies that the eigenvalue of the ground state is $-\frac{1}{4}+o(\gamma^2)$. So we have at $o(\gamma)$
\begin{equation}
\lambda = 2\nu=\frac{2\pi}{\beta}(1-\gamma+o(\gamma^2))=\frac{2\pi}{\beta}\left[1-\frac{\beta\K}{(\beta\J)^2}+o((\tfrac{\beta\K}{(\beta\J)^2})^2)\right]
\end{equation}

\section{Derivation of the chaos exponent from retarded kernel}\label{App:chaos ladder}
The $F(t_1,t_2)$ function satisfies the following equation which comes from a set of ladder diagrams
\begin{equation}\label{chaos_ladder}
F(t_1,t_2)=\int dt_3dt_4 K_R(t_1,t_2;t_3,t_4)F(t_3,t_4)
\end{equation}
The retarded kernel $K_R(t_1,t_2;t_3,t_4)$ is defined by
\begin{align}
K_R(t_1,t_2;t_3,t_4)&=G_R(t_{13})G_R(t_{24})\int_0^\infty d\alpha J^2(\alpha)(\alpha q-1)2^{2-\alpha q}[G_{lr}(t_{34})]^{\alpha q-2}\\
&=\theta(t_{13})\theta(t_{24})\partial_g U(g(\tfrac{\beta}{2}+it_{34})) \label{retarded_kernel}
\end{align}
where $G_R(t)$ is the retarded propagator, which in the large $q$ limit is just $\theta(t)$, and $G_{lr}$ is the Wightman correlator with points separated by half of the thermal circle ($G_{lr}(t)=G(\frac{\beta}{2}+it)$).

To solve \eqref{chaos_ladder}, we use the growth ansatz
\begin{equation}\label{growth_ansatz}
F(t_1,t_2)=e^{\lambda_L(t_1+t_2)/2}f_\lambda(t_{12})
\end{equation}
then the Lyapunov exponent is just the values of $\lambda_L$ such that $f(t)$ is an eigenfunction of $K_R$ with eigenvalue one by solving \eqref{chaos_ladder}. 

By substituting \eqref{retarded_kernel} and \eqref{growth_ansatz} into \eqref{chaos_ladder} and taking derivatives with respect to $t_1$ and $t_2$, we obtain the following equation
\begin{equation}\label{lambda_x}
\left[\frac{\lambda_L^2}{4}-\partial_\sigma^2\right]f_\lambda(\sigma)=\partial_g U(g(\tfrac{\beta}{2}+i\sigma))f_\lambda(\sigma)
\end{equation}
with $\sigma=t_{12}$. The calculation of the Lyapunov exponent becomes the quantum mechanics problem of solving the bound state energy with the potential $\partial_g U(g(\tfrac{\beta}{2}+i\sigma))$.
{}
\bibliographystyle{utphys}
\bibliography{LargeqSYK.bib}

\end{document}